\documentclass{article}
\usepackage{hyperref}
\begin{document}

\markboth{C. A. Brannen}
{The Force of Gravity}

\title{The force of gravity in Schwarzschild and Gullstrand-Painlev\'e
coordinates\footnote{This essay received an ``honorable mention'' in the 2009 Essay Competition of the Gravity Research Foundation.} }

\author{C. A. Brannen \\
720 Road N NE,\\
Moses Lake, WA 98837, USA\\
carl@brannenworks.com}

\maketitle

\begin{abstract}
We derive the exact equations of motion (in Newtonian, $F=ma$, form) for test masses in Schwarzschild and Gullstrand-Painlev\'{e} coordinates. These equations of motion are simpler than the usual geodesic equations obtained from Christoffel tensors in that the affine parameter is eliminated. The various terms can be compared against tests of gravity. In force form, gravity can be interpreted as resulting from a flux of superluminal particles (gravitons). We show that the first order relativistic correction to Newton's gravity results from a two graviton interaction.
\end{abstract}

\section{Gullstrand-Painlev\'{e} Coordinates} \label{sec:GP}

In general relativity, the Schwarzschild solution for a spherically symmetric (non rotating) black hole has been known since 1915. The usual choice of coordinates is the one Karl Schwarzschild used in its discovery, Schwarzschild coordinates, which are characterized as keeping the metric diagonal, but have a coordinate singularity at $r=2M$:
\begin{equation}\label{eq:Schwarz}
(d\tau)^2 = \left(1-2M/r\right)dt^2
- dr^2/(1-2M/r) - r^2(d\theta^2 + \sin^2(\theta)\;d\phi^2),
\end{equation}
where we have chosen coordinates with $G=c=1$. Note that if we multiply $r$ and $t$ by $M$, the metric will end up with an overall multiple of $M^2$ which we can cancel. For convenience, we will do this both with the Schwarzschild and GP coordinates. The reader can reinsert $M$ by making the reverse substitution.

Gullstrand-Painlev\'{e} (GP) coordinates were discovered by Allvar Gullstrand\footnote{Gullstrand had a primary role in denying Einstein a Nobel prize for relativity.} \cite{01Gullstrand} and Paul Painlev\'{e} \cite{02Painleve} in 1921/1922:
\begin{equation}\label{eq:GPcoords}
d\tau^2 = (1-2M/r)dt^2 - 2\sqrt{2M/r}\;dt\;dr - dr^2 - r^2(d\theta^2 + \sin^2(\theta)\;d\phi^2).
\end{equation}
While the metric is not diagonal, the curvature is concentrated into the $dt^2$ and $dr\;dt$ terms. The purely spatial terms of the metric, $-dr^2 - r^2(d\theta^2 + \sin^2(\theta)\;d\phi^2)$, are identical to the spatial part of the natural metric for a flat (Minkowski) space. This makes GP coordinates a natural choice for a model of the gravitational force by an exchanged particle (which we will call the graviton).

A major step towards unifying general relativity with quantum mechanics was made by Anthony Lasenby, Chris Doran, and Stephen Gull in 1993-1998 when they invented ``Gauge Theory Gravity'' (GTG).\cite{03DLG1993,04LDG1998,05HestSGwGC,06HestSGwGC,07HestGTGwGC} In their theory, they rewrote the tensor theory of general relativity into the language of Dirac's gamma matrices \cite{08Heste1981,09Havel2001} (also called, with various subtle shades of meaning, the ``Spacetime Algebra'', ``Geometric Algebra'', or ``Clifford Algebra''). This enables calculations of interactions of fermions with black holes. \cite{10dolan-2006-74,11lasenby-2005-72,12doran-2005-71}

The GTG differed from general relativity only in that it was built on a flat (Minkowski) background space and consequently could not support topologically interesting solutions to Einstein's field equations. Those who might doubt the existence of science fiction topics such as wormholes would find the GTG a substantial improvement over standard general relativity. 

The application of GP coordinates to the GTG is not obvious from their papers; a short explanation may be useful. The metric for a black hole in GTG, requires the definition of four ancillary functions of radius, $f_1$, $f_2$, $g_1$, and $g_2$ and (see equation (52) of
Ref.~\cite{03DLG1993} ) is:
\begin{equation}\label{eq:GTGcoords}
d\tau^2 = (1-2M/r)dt^2 - 2(f_1g_2-f_2g_1)dr\;dt - (f_1^2 - f_2^2)dr^2 - r^2(d\theta^2 + \sin^2(\theta)d\phi^2).
\end{equation}
The four functions of radius define the directional derivatives with respect to $r$ and $t$. One has a choice of gauge for the radial direction that allows one to choose $g_2$ arbitrarily, and from this compute $f_1$, $f_2$, and $g_1$. One requires that $f_1$ and $g_1$ go to one at infinity, while $f_2$ and $g_2$ approach zero. To obtain GP coordinates, one would use the freedom in $g_2$ to require that $f_1^2-f_2^2 = 1$. This gives a metric whose spatial portion is flat and therefore is GP. See Refs.~\cite{13NewForm2000} and \cite{14River2004} for generalizations of GP to the Kerr metric.

\section{The Force of Relativistic Gravity} \label{sec:GPforce}

When computing orbits in general relativity, the easiest and traditional method is to use Christoffel tensors $\Gamma^\mu_{\nu\lambda}$. This gives four differential equations in the affine parameter $q$:
\begin{equation}\label{eq:GeodesicEqn}
\frac{d^2x^\mu}{dq^2} + \Gamma^\mu_{\nu\lambda}\frac{dx^\nu}{dq}\frac{dx^\lambda}{dq} = 0,
\end{equation}
for $\mu = 0, 1, 2$, and $3$. For massive particles, the affine parameter $q$ can be taken to be the proper time but this fails in the massless case. Newtonian gravitation avoids the use of $q$ and so is able to get by with just three differential equations in $t$:
\begin{equation}
m\frac{d^2x^j}{dt^2} + G m \frac{d\Phi}{dx^j} = 0,
\end{equation}
that is, $m\vec{a} - \vec{F} = 0$.

To find the Newtonian equations of motion from the Schwarzschild metric, the most direct way is to first note that geodesic paths extremize $s = \int ds$. So one can write $ds$ in terms of $dx/dt$, $dy/dt$, $dz/dt$ and use the Euler-Lagrange equations to vary $\int ds$. Following the Newtonian tradition, we will abbreviate $dx/dt$ by $\dot{x}$ and similarly for $\ddot{x}$, etc. The result\footnote{The author used the symbolic computation software MAXIMA. A Java applet that demonstrates various properties of relativistic gravity that uses these formulas is at \href{http://www.brannenworks.com/Gravity/index.html}{www.brannenworks.com/Gravity}} for Schwarzschild coordinates is a few terms in $r$ and $r-2$:
\begin{equation}\label{eq:SchwarzEqn}
\begin{array}{rcl}
\ddot{x}\;r^4(r-2) &=& 3xr^2\dot{r}^2 - x(r-2)^2 + 4y\dot{r}(y\dot{x}-x\dot{y})+
4z\dot{r}(z\dot{x}-x\dot{z})\;+\\
&&2r(r-2)\dot{y}(y\dot{x}-x\dot{y})+2r(r-2)\dot{z}(z\dot{x}-x\dot{z}),
\end{array}
\end{equation}
and similarly for $\ddot{y}$ and $\ddot{z}$. GP coordinates are simpler in that they need only powers of $r$:
\begin{equation}\label{eq:PainleveEqn}
\begin{array}{rcl}
\ddot{x}r^5 &=& -xr^2+2xr-2r^2\dot{y}(x\dot{y}-y\dot{x})-2r^2\dot{z}(x\dot{z}-z\dot{x})+3xr^2\dot{r}^2+\\
&&\sqrt{2r}(3r^2\dot{x} - r^3\dot{x}|\dot{\vec{r}}|^2 + 1.5r^3\dot{x}\dot{r}^2
+ 2y(x\dot{y}-y\dot{x}) + 2z(x\dot{z}-z\dot{x}))
\end{array}
\end{equation}
The easiest way to verify the above equations is to choose a set of random positions and velocities, and compare the acceleration with that computed from the geodesic equations. One finds that the above are exact, and that they work for massless as well as massive test particles.

We can modify general relativity by making changes to the various terms in the above. This gives us information on which terms have been fixed by experimentaal tests. For example, the terms contribute to the small deflection of starlight (DoS) and perihelion advance of Mercury (PoM) in proportion as follows:
\begin{equation}\label{eq:RadialTests}
\begin{array}{cccc}
\textrm{GP term}&\textrm{DoS}&\textrm{PoM}\\ \hline
-x/r^3                                   &+1/2& 0  &\\
2x/r^4                                   &0   &-1/3&\\
-2\dot{y}(x\dot{y}-y\dot{x})/r^3         &+1  &+4/3&\\
3x\dot{r}^2/r^3                          &-1/2&0   &\\
3\sqrt{2}\dot{x}/r^{2.5}                 &0   &0   &\\
-\sqrt{2}\dot{x}|\dot{\vec{r}}|^2/r^{1.5}&0   &0   &\\
1.5\sqrt{2}\dot{x}\dot{r}^2/r^{1.5}      &0   &0   &\\
\sqrt{8}y(x\dot{y}-y\dot{x})/r^{4.5}     &0   &0   &\\ \hline
\textrm{Total:}                          &+1  &+1  &
\end{array}
\end{equation}
Only the first four terms contribute to the solar system tests of general relativity.

A characteristic of a spherically symmetric gravitational force is that it depends only on the distance to the body $r$, and two velocities, the raidal velocity $\dot{r}$ and the velocity perpendicular $\dot{h}$. For the Schwarzschild coordinates, we find:
\begin{equation}
\begin{array}{rcl}
\ddot{r} &=& (-1 -2\dot{h}^2)/r^2 + 3\dot{r}^2/(r(r-2)) + 2/r^3,\\
\ddot{h} &=& 2\dot{r}\dot{h}/r^2 + 4\dot{r}\dot{h}/(r^2(r-2)).
\end{array}
\end{equation}
The $-1/r^2$ is the Newtonian part of the force. Other than the $+2/r^3$, the remaining terms are all of 2nd order in $\beta =\sqrt{\dot{x}^2+\dot{y}^2}$.

For GP coordinates, grouping terms by their order in $r$ we find:
\begin{equation}
\begin{array}{rcl}
\ddot{r} &=&  -\sqrt{2}\dot{r}\dot{h}^2/r^{1.5} + (-1 -2\dot{h}^2 + 3\dot{r}^2)/r^2
+ 3\sqrt{2}\dot{r}/r^{2.5} + 2/r^3,\\
\ddot{h} &=& -\sqrt{2}\dot{h}^3/r^{1.5} + 2\dot{r}\dot{h}/r^2 +\sqrt{2}\dot{h}/r^{2.5}
+\sqrt{1/2}\dot{h}\dot{r}^2/r^{4.5}.
\end{array}
\end{equation}
For $\dot{r} = \dot{h} = 0$, Schwarzschild and Painleve coordinates give the same ``static'' acceleration:
\begin{equation}\label{eq:gravitostatics}
\begin{array}{rcl}
\ddot{r} &=& -1/r^2 +2/r^3,\\
\ddot{h} &=& 0.
\end{array}
\end{equation}
This acceleration differs from the Newtonian acceleration by the addition of the $+2/r^3$ term. In particular, the static acceleration is zero at the event horizon, is negative inside the event horizon, and is everywhere identical for both GP and Schwarzschild coordinates.

\section{Gravitons in the Gravitostatic Limit} \label{sec:Flux}

The classical electric force between charged bodies follows an exact $1/r^2$ law. This can be attributed to the electric force being carried by an elementary particle (the photon) that is massless and whose intensity naturally decreases according to the area over which the photons are spread. In making these sorts of arguments, we must recognize that we cannot assume that the velocity of the gravitons involved is the same as the speed of light. In particular, in GP coordinates, a particle falling through the event horizon exceeds the speed of light (that is, $|\dot{r}| > 1$).

Gravitons capable of producing such a force must also exceed the speed of light. Such a theory would have gravity waves also travel faster than light. The experimental measurement of the speed of gravity is a subject much discussed in the physics literature. The mainstream view is that the speed has not yet been measured.\cite{15will-2005-3} Consequently, our discussion, while speculative, is not yet eliminated by experiment.

Let the number of gravitons passing through a sphere of radius $r$ be $N(r)$. From Eq.~(\ref{eq:gravitostatics}), we have that
\begin{equation}\label{eq:Nofr}
N(r) \propto |4\pi r^2 (-1/r^2+2/r^3)| = 4\pi(1 - 2/r).
\end{equation}
As the radius increases, $N(r)$ increases until it approaches a constant corresponding to Newtonian gravitation. We will use $n$ for this constant so that $N(r) = (1-2/r)n$. The number of gravitons per unit volume at a distance $r$ from the gravitating body is
\begin{equation}\label{eq:rhoOfR}
\rho(r) = N(r)/(4\pi r^2) = (1/r^2-2/r^3)n/(4\pi).
\end{equation}
For large $r$, this approaches $n/(4\pi r^2)$.

The number of gravitons appearing at a radius of $r$ is given by the derivative of $N(r)$:
\begin{equation}
N'(r) = 2n/r^2.
\end{equation}
Dividing this number by the surface area over which these gravitons appear, $4\pi r^2$, gives the rate of change of gravitons per unit volume:
\begin{equation}\label{eq:RhoDotOfR}
\rho'(r) = N'(r)/(4\pi r^2) = 2n/(4\pi r^4).
\end{equation}
The number of gravitons appearing at the radius $r$, Eq.~(\ref{eq:RhoDotOfR}), is proportional to $1/r^4$. This is approximately proportional to the square of the graviton density at that radius, Eq.~(\ref{eq:rhoOfR}). Consequently, we obtain that the graviton density increases at a rate proportional to the square of the density of gravitons. In a perturbation theory of gravitons, this term would arise by allowing two gravitons to interact to create three (or more) all moving in the same direction. That is, in addition to their emission from massive bodies, gravitons are emitted by stimulated emission somewhat similar to the effect in lasers.


\section*{Acknowledgments}
The financial assistance of many people was necessary to provide the author the time to explore these ideas, in particular, Mark Mollo, Liquafaction Corporation, its employees, and those who have done business with it, especially Lloyd Haisch; also the author's parents and a close friend who wishes to remain unnamed.


\end{document}